\input harvmac
\parskip 7pt plus 1pt
\noblackbox

\input epsf

\def\la{\mathrel{\mathpalette\fun <}}
\def\ga{\mathrel{\mathpalette\fun >}}
\def\fun#1#2{\lower3.6pt\vbox{\baselineskip0pt\lineskip.9pt
  \ialign{$\mathsurround=0pt#1\hfil##\hfil$\crcr#2\crcr\sim\crcr}}}
\relax


\lref\lrreview{D.~H.~Lyth and A.~Riotto,
Phys.\ Rept.\  {\bf 314} (1999) 1 
[hep-ph/9807278].}

\lref\guth{A.~H.~Guth,
 Phys.\ Rev.\ D {\bf 23} (1981) 347.}

\lref\birdav{N.~D.~Birrell and P.~C.~W.~Davies,
{\it Quantum Fields In Curved Space}, Cambridge University Press, 1982.}

\lref\pngb{K.~Dimopoulos, D.~H.~Lyth, A.~Notari and A.~Riotto, 
JHEP {\bf 0307}
(2003)
053 
[hep-ph/0304050].}



\lref\turner{For a recent review, see 
M.S. Turner,
Phys.\ Rept.\ {\bf 333} (2000) 619.}


\lref\griestkam{K. Griest and M. Kamionkowski,
Phys. Rev. Lett. {\bf 64} (1990) 615.}

\lref\mc{J.~McDonald,
Phys.\ Lett.\ B {\bf 511} (2001) 1
[hep-ph/0011198].}


\lref\ad{T.~Asaka, K.~Hamaguchi, M.~Kawasaki and T.~Yanagida,
Phys.\ Lett.\ B {\bf 464} (1999) 12
[hep-ph/9906366];
R.~Allahverdi and M.~Drees,
Phys.\ Rev.\ Lett.\  {\bf 89} (2002) 091302
[hep-ph/0203118].}

\lref\da{R.~Allahverdi and M.~Drees,
Phys.\ Rev.\ D {\bf 66} (2002) 063513
[hep-ph/0205246].}

\lref\sel{U.~Seljak {\it et al.},
astro-ph/0407372.}


\lref\ckra{D.~J.~Chung, E.~W.~Kolb and A.~Riotto,
Phys.\ Rev.\ D {\bf 59} (1999) 023501 
[hep-ph/9802238].}

\lref\ckrb{D.~J.~Chung, E.~W.~Kolb and A.~Riotto,
Phys.\ Rev.\ Lett.\ {\bf 81} (1998) 4048 
[hep-ph/9805473].}

\lref\ckrc{D.~J.~Chung, E.~W.~Kolb and A.~Riotto,
Phys.\ Rev.\ D {\bf 60} (1999) 063504 
[hep-ph/9809453].}

\lref\ckrd{E.~W.~Kolb, D.~J.~Chung and A.~Riotto,
hep-ph/9810361.}

\lref\ckre{D.~J.~H.~Chung, P.~Crotty, E.~W.~Kolb and A.~Riotto,
Phys.\ Rev.\ D {\bf 64} (2001) 043503
[hep-ph/0104100].}

\lref\kuzmin{V.~Kuzmin and I.~Tkachev,
JETP Lett.\ {\bf 68} (1998)  271 
[hep-ph/9802304].}


\lref\mr{A.~Masiero and A.~Riotto,
Phys.\ Lett.\ {\bf B289} (1992)  73 
[hep-ph/9206212].}

\lref\Kolba{E.~W.~Kolb, A. Linde and A.~Riotto, 
Phys. Rev. Lett. 
{\bf 77} (1996)  4290 [hep-ph/9606260];
E.~W.~Kolb, A.~Riotto and I.~I.~Tkachev,
Phys.\ Lett.\ B {\bf 423} (1998)  348
[hep-ph/9801306].}

\lref\Giudicea{G.~F.~Giudice, M.~Peloso, A.~Riotto and I.~Tkachev,
JHEP {\bf 9908} (1999)  014 
[hep-ph/9905242].}

\lref\tonireview{A.~Riotto,
``Inflation and the theory of cosmological perturbations,''
Lectures given at ICTP Summer School on Astroparticle Physics and
Cosmology, Trieste, Italy, 17 Jun - 5 Jul 2002.
Published in Trieste 2002, Astroparticle physics and cosmology 317-413 
[hep-ph/0210162].}

\lref\lindebook{A.D. Linde, {\it Physics and Inflationary 
Cosmology}, Harwood, Chur, Switzerland, 1990.}

\lref\hybrid{A.~D.~Linde, 
Phys.\ Rev.\ D {\bf 49} (1994)
748 [astro-ph/9307002].}

\lref\prz{L.~Pilo, A.~Riotto and A.~Zaffaroni,
hep-th/0401004.}


\lref\curvaton{S.~Mollerach,
Phys.\ Rev.\ D {\bf 42} (1990) 313; 
A.~D.~Linde and V.~Mukhanov,
Phys.\ Rev.\ D {\bf 56} (1997) 535;
K.~Enqvist and M.~S.~Sloth,
Nucl.\ Phys.\ B {\bf 626} (2002) 395 
[hep-ph/0109214];
T.~Moroi and T.~Takahashi,
Phys.\ Lett.\ B {\bf 522} (2001) 215
[Erratum-ibid.\ B {\bf 539} (2002) 303]
[hep-ph/0110096];
D.~H.~Lyth and D.~Wands,
Phys.\ Lett.\ B {\bf 524} (2002)  5 
[hep-ph/0110002].}

\lref\GD{O.~DeWolfe and S.~B.~Giddings,
Phys.\ Rev.\ D {\bf 67} (2003)  066008 
[hep-th/0208123].}

\lref\GKP{S. Giddings, S. Kachru and J. Polchinski, 
Phys. Rev. D {\bf 66}
(2002) 106006 [hep-th/0105097].}

\lref\rsI{L.~Randall and R.~Sundrum,
Phys.\ Rev.\ Lett.\  {\bf 83} (1999) 3370
[hep-ph/9905221].}



\lref\KS{I. Klebanov and M.J. Strassler, 
JHEP {\bf 0008} (2000) 052
[hep-th/0007191].}

\lref\KKLMMT{
S.~Kachru, R.~Kallosh, A.~Linde, J.~Maldacena, L.~McAllister and 
S.~P.~Trivedi,
JCAP {\bf 0310} (2003)  013
[hep-th/0308055].}

\lref\KKLT{S.~Kachru, R.~Kallosh, A.~Linde and S.~P.~Trivedi,
Phys.\ Rev.\ D {\bf 68} (2003) 046005
[hep-th/0301240].}

\lref\KKLMMTrefs{
J.~P.~Hsu, R.~Kallosh and S.~Prokushkin, 
JCAP {\bf 0312} (2003) 009
[hep-th/0311077];
F.~Koyama, Y.~Tachikawa and T.~Watari, 
Phys.\ Rev.\ D {\bf 69} (2004) 106001
[hep-th/0311191];
H.~Firouzjahi and S.~H.~H.~Tye, 
Phys.\ Lett.\ B {\bf 584} (2004) 147
[hep-th/0312020];
J.~P.~Hsu and R.~Kallosh, 
JHEP {\bf 0404}
(2004) 042 [hep-th/0402047];
C.~P.~Burgess, J.~M.~Cline, H.~Stoica and F.~Quevedo,
hep-th/0403119;
O.~DeWolfe, S.~Kachru and H.~Verlinde, 
JHEP {\bf 0405} (2004) 017 [hep-th/0403123];
N.~Iizuka and S.~P.~Trivedi, 
hep-th/0403203.}

\lref\DD{F.~Denef, M.~R.~Douglas and B.~Florea,
JHEP {\bf 0406} (2004) 034
[hep-th/0404257].}

\lref\giant{O.~DeWolfe, S.~Kachru and H.~Verlinde,
JHEP {\bf 0405} (2004) 017 
[hep-th/0403123].}

\lref\igor{V.~Kuzmin and I.~Tkachev,
Phys.\ Rev.\ D {\bf 59} (1999) 123006
[hep-ph/9809547].}

\lref\braneinflation{S. Alexander, 
Phys. Rev. D {\bf 65} (2002) 023507
[hep-th/0105032]; G. Dvali, Q. Shafi and S. Solganik, 
hep-th/0105203; C.P. Burgess, M. Majumdar, D.
Nolte, F. Quevedo, G. Rajesh and R.J. Zhang, 
JHEP {\bf 07} (2001) 047
[hep-th/0105204]; G. Shiu and S.H. Tye, 
Phys. Lett. B {\bf 516} (2001) 421 [hep-th/0106274];
D. Choudhury, D. Ghoshal, D.P. Jatkar,
S. Panda, 
JCAP {\bf 0307} (2003) 009
[hep-th/0305104].}

\lref\lepto{M.~Fukugita and T.~Yanagida,
Phys.\ Lett.\ B {\bf 174} (1986) 45.}

\lref\leptonoi{G.~F.~Giudice, A.~Notari, M.~Raidal, A.~Riotto and A.~Strumia,
Nucl.\ Phys.\ B {\bf 685} (2004) 89
[hep-ph/0310123].}

\lref\softl{Y.~Grossman, T.~Kashti, Y.~Nir and E.~Roulet,
Phys.\ Rev.\ Lett.\  {\bf 91} (2003) 251801
[hep-ph/0307081];
G.~D'Ambrosio, G.~F.~Giudice and M.~Raidal,
Phys.\ Lett.\ B {\bf 575} (2003) 75
[hep-ph/0308031].}

\lref\pilaf{A.~Pilaftsis and T.~E.~J.~Underwood,
Nucl.\ Phys.\ B {\bf 692} (2004) 303
[hep-ph/0309342].}

\lref\sndom{K.~Hamaguchi, H.~Murayama and T.~Yanagida,
Phys.\ Rev.\ D {\bf 65} (2002) 043512 
[hep-ph/0109030].}

\lref\allah{R.~Allahverdi and M.~Drees,
Phys.\ Rev.\ D {\bf 69} (2004) 103522
[hep-ph/0401054].}

\lref\infa{K.~Kumekawa, T.~Moroi and T.~Yanagida,
Prog.\ Theor.\ Phys.\  {\bf 92} (1994) 437
[hep-ph/9405337].}

\lref\bound{S.~Davidson and A.~Ibarra,
Phys.\ Lett.\ B {\bf 535} (2002) 25 
[hep-ph/0202239].}

\lref\strum{T.~Hambye, Y.~Lin, A.~Notari, M.~Papucci and A.~Strumia,
Nucl.\ Phys.\ B {\bf 695} (2004) 169
[hep-ph/0312203].}

\lref\kawa{M.~Kawasaki, K.~Kohri and T.~Moroi,
astro-ph/0402490.}

\lref\llbook{A.~R.~Liddle and D.~H.~Lyth,
{\it Cosmological inflation and large-scale structure},
 Cambridge University Press, 2000.}

\hskip 1cm
\vskip 0.2in

\Title{\vbox{\baselineskip12pt \hbox{CERN-PH-TH/2004-151}
\hbox{hep-ph/0408155} }}
{\vbox{\centerline{Heavy Particles from Inflation}}}
\vskip .2in
\centerline{G.F. Giudice$^{a}$,
A. Riotto$^{b}$, A. Zaffaroni$^{c}$}
\vskip .2in
\centerline{$^a$
\it CERN Theory Division, Geneva, Switzerland}
\vskip .1in
\centerline{$^b$ \it INFN Sezione di Padova, 
Padua, Italy}
\vskip .1in
\centerline{$^c$ \it Universit\`a di Milano-Bicocca, Milan, 
Italy}

\vskip .2in
\baselineskip18pt 
\noindent
We describe a simple and efficient mechanism by which very heavy 
particles are copiously created from a primordial inflationary epoch. 
It works for scalar fields which are massless or very light during 
inflation and acquire a large mass right after the end of inflation. 
Such particles can exist in realistic scenarios, as we illustrate
with several examples in the context of both field and string theory.
Long-wavelength fluctuations of these fields are generated during 
inflation with an almost scale-invariant spectrum and may give the 
dominant contribution to the energy density of the heavy fields at 
late times. Applications of our results to superheavy dark matter 
and leptogenesis are discussed.

\Date{August 2004}

\eject
\baselineskip20pt 

\newsec{Introduction}

Inflation has become the standard paradigm for explaining the
homogeneity and the isotropy of our observed Universe \guth\
 (for a recent review, see ref.~\lrreview).
If at  some primordial epoch  the energy density of the 
Universe is dominated by some scalar field -- 
the inflaton -- whose kinetic energy is negligible, the corresponding vacuum
energy gives rise to an exponential growth of the scale factor. 
During this phase a small, smooth region of size of the order
of the Hubble radius grew so large that it easily encompasses
the comoving volume of the entire presently observed Universe
and one can understand why the observed Universe is  homogeneous and
isotropic to such high accuracy. Furthermore, 
it is now clear that structure in the Universe
comes primarily from an almost scale-invariant superhorizon curvature
perturbation. This perturbation originates presumably from the vacuum 
fluctuation,
 during the almost-exponential inflation,  of some field with mass
much less than the Hubble parameter $H$ during inflation. Indeed, 
any scalar field whose mass is lighter than $H$ 
suffers, in a (quasi) de Sitter epoch, quantum fluctuations whose
power spectrum is indepedent of the length scales,
if they are superhorizon \refs{\lindebook,\llbook,\tonireview}. 
On the contrary, 
 quantum fluctuations of 
scalar fields whose mass $M$ is larger  than the Hubble rate 
during inflation
are not efficiently excited, and their power spectrum is suppressed
as $\sim e^{-2 M^2/H^2}$ \birdav . 

While massive fields may be
produced after inflation, {\it e.g.} during the process of preheating \Kolba ,
in this paper we show that there are interesting cases where 
fluctuations 
generated during inflation rather than quantum fluctuations produced
after inflation give the dominant contribution to heavy particle production.
The mechanism of heavy particle production from inflation we have in mind is 
based on a simple observation:  particles which are
massive in the present-day vacuum  could have been very light
during inflation. This implies that fluctuations of 
a generic scalar field, with mass $M \ll H$ during inflation and
$M \gg H$ right after inflation,  
are copiously generated during inflation with an almost
scale-invariant spectrum and become  heavy and  non-relativistic right at the
end of inflation. 
This provides an efficient mechanism to create massive
particles from inflation without any exponential suppression.
We will show that the number density of heavy particles at late times
comes from the long-wavelength modes which are far outside the horizon at the
end of inflation. As long as they are outside the horizon, these
modes do not have a truly particle nature, but nevertheless may provide the
dominant contribution to the energy density stored in the heavy 
field. 

In sect.~\S2\ we study the quantum creation of heavy particles
from inflation. Next, we describe
some interesting cases in which
scalar fields can be massive in the present vacuum, but
indeed massless
during inflation: field-theoretical models are discussed in sect.~\S3\ and
D-brane models in sect.~\S4 . 
We
devote sect.~\S5\
to possible applications of our findings.
Finally, sect.~\S6\ contains our conclusions.

\newsec{Heavy particles from inflation}

We start by considering a
field $\chi$ which is nearly massless ($M < H$)
during inflation, but very massive ($M\gg H$) right at the end
of inflation. Our goal is to compute its abundance
at the end of the inflationary stage and to
show that the production of such massive field is very efficient.

Since the field $\chi$ is nearly
massless during the inflationary epoch, its vacuum fluctuations
in momentum space $\delta\chi_k$ 
are generated during the almost-exponential inflation. Indeed, 
the perturbations $\delta\chi_k$
acquire a nearly scale-invariant 
spectrum \refs{\lrreview,\lindebook,\llbook,\tonireview}.
The fluctuations of the $\chi$ field have exponentially large
wavelengths and -- as long as their wavelength is larger than the horizon --
for practical purposes they behave
as a homogeneous classical field. 

During inflation,
the fluctuations of the $\chi$ field
obey the equation
\eqn\per{\delta\ddot{\chi}_k+3\,H\, \delta\dot{\chi}_k+
\left({k\over a}\right)^2\delta\chi_k=0, }
where $a$ is the scale factor and $H$ is the Hubble rate. They  can 
can be conveniently described in terms of the variance
\eqn\variance{\langle \chi^2\rangle=\int\,{d^3k\over (2\pi)^3}\,
\left|\delta\chi_k\right|^2,}
 which, during inflation, obeys the equation \lindebook\
\eqn\growth{{d\langle \chi^2\rangle\over dt}={H^3\over 4 \pi^2}.}
In slow-roll models of inflation
\refs{\lrreview,\lindebook,\llbook,\tonireview}, 
the Hubble rate is not exactly constant, but
slowly decreases with time according to
the relation
\eqn\rateh{\dot H=-\epsilon\,H^2,}
where $\epsilon=[V^\prime(\phi)/ V(\phi)]^2/(16\pi G)$
is one of the  slow-roll parameters. 
Combining equations \growth\ and \rateh\ and assuming
that $\epsilon$ does not change appreciably 
with  time (in slow-roll, time derivatives of 
slow-roll paramaters are proportional to the  squares of  slow-roll
parameters and can be legally neglected), one finds
\eqn\a{{d\langle \chi^2\rangle\over dt}=-{1\over 8\, \epsilon\, \pi^2}\,{
d H^2\over dt}.}
If we indicate by $H_*$ the initial value of the Hubble rate at the beginning
of inflation, the value of the fluctuations of the massless
field $\chi$ depends on such an initial value as 
\eqn\result{\langle \chi^2\rangle={H_*^2\over 8\,\epsilon\,\pi^2}.}
This means that the largest contribution to the variance of the $\chi$ field is given
by the time when the Hubble constant took its largest value 
or, equivalently,
that the number density of $\chi$ particles at late times comes 
predominantly from
those infra-red long-wavelength modes which are far outside 
the horizon after inflation and
crossed the horizon much before the end of inflation. 
As long as they remain outside the horizon these modes
do not manifest a truly particle-like behaviour, but are 
indistinguishable from a classical
homogenous field whose amplitude at the end of inflation
is $\sim\langle \chi^2\rangle^{1/2}\sim 10^{-1}\,(H_*/\epsilon^{1/2})$.
This amplitude depends upon the model of inflation, but one can perform 
a simple estimate
of it by assuming that the Universe expanded for at least $\sim$ 
60 $e$-folds before
the end of inflation in order
to solve the horizon and the flatness problems. In many models of inflation
(but not in all) the slow-roll parameter $\epsilon\sim 1/N$,
when there are $N$ $e$-folds to go till the end of inflation. 
Taking $N\sim 60$, one
finds that the amplitude of the variance of the field $\chi$ at the end of
inflation is given by
\eqn\estimate{\langle \chi^2\rangle^{1/2}\sim 0.9\, H_* ,}
though inflationary models typically
provide a number of $e$-folds much larger than  $ 60$
and the fluctuations of the $\chi$ field at the end of inflation are 
expected to  be greater. 

In models of inflation
where the inflaton field is well anchored at its false vacuum, such as the
model of stringy old inflation proposed in  ref.~\prz , the Hubble rate is
 constant and the variance 
of the $\chi$-fluctuations is given by 
$\langle \chi^2\rangle={H_*^2\over 4\,\pi^2}\, N$ thus providing 
a quantitative estimate similar to eq.~\estimate .

Suppose now that the field $\chi$ acquires a large mass $M$ right at the end
of inflation. This large mass is supposed to be greater than the Hubble rate
at the end of inflation and at the beginning of the reheating stage.
The long wavelength modes satisfy the equation
\eqn\perb{\delta\ddot{\chi}_k+3\,H\, \delta\dot{\chi}_k+
M^2\delta\chi_k\simeq 0, }
which tells us that modes beyond the horizon scale like
 $\delta\chi_k\propto a^{-3/2}\,\cos\,Mt$.
This means that the long wavelength modes of $\chi$ become nonrelativistic and act like
a classical homogeneous field. Their associated number
density will be simply given by 
\eqn\n{n_\chi={\rho_\chi\over M}={M\langle\chi^2\rangle\over 2 }.}
The ratio at the
end of inflation between the energy density stored in the heavy $\chi$ particles
and the energy density stored in the inflaton field is
\eqn\rr{{\rho_\chi\over \rho_\phi}\simeq {1\over 6\pi \epsilon}\,
\left({M\over M_p}\right)^2 ,}
where $M_p=1.2\times 10^{19}$~GeV is the Planck mass.

The energy density in the heavy nonrelativistic particles
scales as $a^{-3}$, where $a$ is the scale factor. Meanwhile, during reheating
the energy density stored in the inflaton field $\phi$ is redshifted away as $a^{-3}$
as well. This means that the initial ratio, 
given by eq.~\rr ,
does not change with time
until the time when reheating occurs. At the time of reheating, the energy
density stored in the inflaton oscillations gets converted into
radiation with entropy density $s=2\pi^2 g_* T_r^3/45$, where $g_*$ is the
number of light degrees of freedom and $T_r$ is the reheating temperature.
This yields
\eqn\ratio{{n_\chi\over s}={3T_r\over 4M}\, {\rho_\chi\over \rho_\phi}\simeq
{M\, T_r\over 8\pi \epsilon\, M_p^2}.}
 
This ratio can be sizeable for large values of masses $M$ and small values
of the slow-roll parameter $\epsilon$. The production of heavy
particles which are massless during inflation and acquire
a mass right after inflation
can therefore 
be rather efficient. 

Equation \ratio\ is approximately valid also in the case of a particle
with the same 
mass $M$ ($\la H$) before and after inflation (see also ref.~\mc). 
In this case,
using the condition
that $\chi$ is not thermally produced ($T_r\la M$), we obtain an upper
bound on its abundance, $n_\chi /s \la H^2/(8\pi \epsilon M_p^2)$. The present
constraint on tensor perturbations ($H< 2\times 10^{14}$~GeV \sel), therefore
requires $n_\chi /s \la 1\times 10^{-11}/\epsilon$. In the case under 
consideration in this paper, this limit can be violated, 
making use of the property that
the mass
$M$ after inflation can actually be much larger than $H$.

It is also interesting to compare the result of eq.~\ratio ,
obtained for minimally-coupled scalars massless during inflation, with the 
case of conformally-coupled scalar particles or of fermions $\psi$, which
have the same mass
before and after inflation.
Now the mass term is the only
source of conformal-symmetry breaking, and such particles $\psi$ can be
generated during the 
non-adiabatic transition
between the end of inflation and the beginning of the radiation
phase. 
Numerical simulations in chaotic inflation
show that \igor\ 
$\rho_\psi /\rho_\phi \simeq 10^{-3} HM_\psi /M_p^2$ 
when $M_\psi\la H$, while $\rho_\psi /\rho_\phi$
is exponentially suppressed for $M_\psi \gg H$.
This gives 
\eqn\pip{{n_\psi \over s}\simeq 10^{-3} {HT_r \over M_p^2}.}
Again, the requirement that $\psi$ is not thermally produced ($T_r\la M$) 
gives a stringent
bound on its relic density $n_\psi /s \la 3\times
10^{-13}(H/2\times 10^{14}~{\rm GeV})^2$.

\newsec{Massless states during inflation which become  massive after inflation}

In this section we want to show that
it is indeed conceivable that states which are massive in the present-day
vacuum could  have been massless, or at least very light, during inflation.
Let us suppose that inflation is driven by a scalar field $\phi$ with
potential $V(\phi)$. There are various possibilities
for which  a given field $\chi$ is massless during inflation and very massive
right at the end of inflation. We describe some of them, but the
list is certainly not exhaustive.

The first option one can envisage is that
the mass $M$ of the $\chi$ field is provided by an interaction term with the
inflaton field itself of the form $g^2\phi^2\chi^2$, where $g$ is some
coupling constant. If during inflation the vacuum expectation value of the inflaton field
is close to the origin in field space, $\phi\simeq 0$, the $\chi$ field
is effectively massless. However, as soon as inflation ends and the
inflaton field rolls down the potential and acquires a large vacuum expectation
$v$, the field $\chi$ becomes massive with mass squared $M^2=g^2 v^2\gg H^2$. This
may happen, for instance, in the so-called small-field models \refs{\lrreview,
\tonireview} where the
potential of the inflaton field, expanded around the origin, is given by
\eqn\pot{V(\phi)=V_0-{m^2\over 2}\phi^2+\cdots .}
Here $V_0$ represents the vacuum energy density driving inflation and the inflaton
field slowly rolls down during inflation with a mass squared 
$m^2\ll H^2\sim V_0/M_p^2$. 
During inflation,
the value of the inflaton field is exponentially close to the origin \lrreview ,
$\phi\sim \phi_{f}\,e^{-N (V_0/m^2 M_p^2)}$, where $N$ is the number of 
$e$-folds
to go till the end of inflation and $\phi_f$ is the value of the inflaton
field at the end of inflation. This means that the field $\chi$ is
practically massless during inflation and its vacuum flucutations can be
generated during the almost-exponential expansion epoch,
as described in sect.~\S2 .

Another possibility may arise in 
hybrid models of inflation \hybrid . The potential relevant for the
inflationary trajectory is given by
\eqn\hyb{V(\phi,S)=\lambda \left(|S|^2-\Lambda^2\right)^2+{m^2\over 2}\phi^2
+h^2 |S|^2\phi^2 .}
The true vacuum of such a potential is given at $|S|=\Lambda$ and 
$\phi=0$. However, if $h^2\phi^2\gg 2\lambda \Lambda^2$, the curvature
of the potential along the $S$-direction becomes positive and the potential reduces
to $V=\lambda \Lambda^4 +{m^2\over 2}\phi^2$. Inflation is driven by the
vacuum energy density $V_0=\lambda \Lambda^4$ and ends when the inflaton field
$\phi$ reaches the critical point $\phi_c^2=2(\lambda/h^2)\Lambda^2$. 
If the particle $\chi$ acquires mass only through couplings with $S$,
it remains massless during inflation, when the vacuum expectation value 
of $S$ vanishes, but it becomes massive after the field $S$ has relaxed
to its true minimum.

The mechanisms described above rely on tree-level potentials, but large masses
for the scalar particles can be induced by radiative corrections. This
is just another incarnation of the usual hierarchy problem of the Higgs
potential, whose solution is still unknown. Supersymmetry or non-linearly
realized symmetries could protect the scalar mass during inflation. A more
exotic possibility is that $\chi$ is the extra component of gauge fields
in theories with more than four dimensions. If the corresponding gauge symmetry
is unbroken during inflation, a mass term for $\chi$ 
(which is a scalar field under the 4-dimensional Lorentz group) is forbidden.
Once gauge symmetry is broken, after inflation, the $\chi$ field
becomes massive.

A second source of concern comes from induced interaction terms of the form
$-(\xi /2)\sqrt{-g}R(g)|\chi|^2$, where $R(g)$ is the scalar curvature,
function of the metrics $g$,
and $\xi$ is a coupling constant. During 
inflation, this term generates a squared mass for $\chi$ equal to $6\xi H^2$,
spoiling the condition $M\ll H$.
This problem is actually generic
in supergravity, since supersymmetric scalars
are conformally coupled ($\xi =1/6$). 
The $H^2$ terms in the $\chi$ mass can be made small, but at the price of
a certain conspiracy between the different supersymmetry-breaking 
contributions.
On the other hand, Goldstone bosons are minimally
coupled ($\xi =0$), and the non-linearly realized symmetry 
can simultaneously protect
the $\chi$ mass from both 
quantum and $H^2$ corrections. This can be realized in 
our scenario if the inflaton background field breaks an exact (or approximate) 
global symmetry:
a Goldstone (or pseudo-Goldstone) 
boson appears during inflation. If the inflaton field vanishes
after inflation, the symmetry is restored and $\chi$ acquires a large mass.
 
Let us close this section with a final comment. 
In sect.~\S2\ we have assumed that the classical field $\chi$ vanishes
during inflation. There is no strong justification for this
assumption in theories where the mass of the field $\chi$ is smaller
than $H$ during inflation. However, the main purpose of the paper is to 
demonstrate that a large contribution to the energy stored
in the heavy field $\chi$ may come from the quantum fluctuations created
during inflation. In this respect, the computation of the energy density
stored in the field $\chi$ presented in sect.~\S2\
should be regarded as conservative.

\newsec{Massless particles in $D$-brane models}

Other set-ups where massless fields during inflation become massive
after the inflationary epochs may arise in models of inflation
inspired by supergravity and (super)string theories. 
In scenarios with $D$-brane inflation, light fields may arise
as pseudo-Goldstone bosons associated to broken translational invariance
\braneinflation .
In these models, the distance between a $D$-brane and an anti $D$-brane
plays the role of the
inflaton. If the branes are embedded in a string compactification
where the background geometry has exact or approximate isometries, 
brane fluctuations transverse to the mutual distance may give rise to
almost massless fields. Inflation ends when the branes come close together
and annihilate. In this process the light fields become very massive
and decay. Indeed, after the end of the inflation the branes have 
disappeared and certainly cannot fluctuate anymore.

\midinsert
\centerline{\epsfxsize=5in\epsfbox{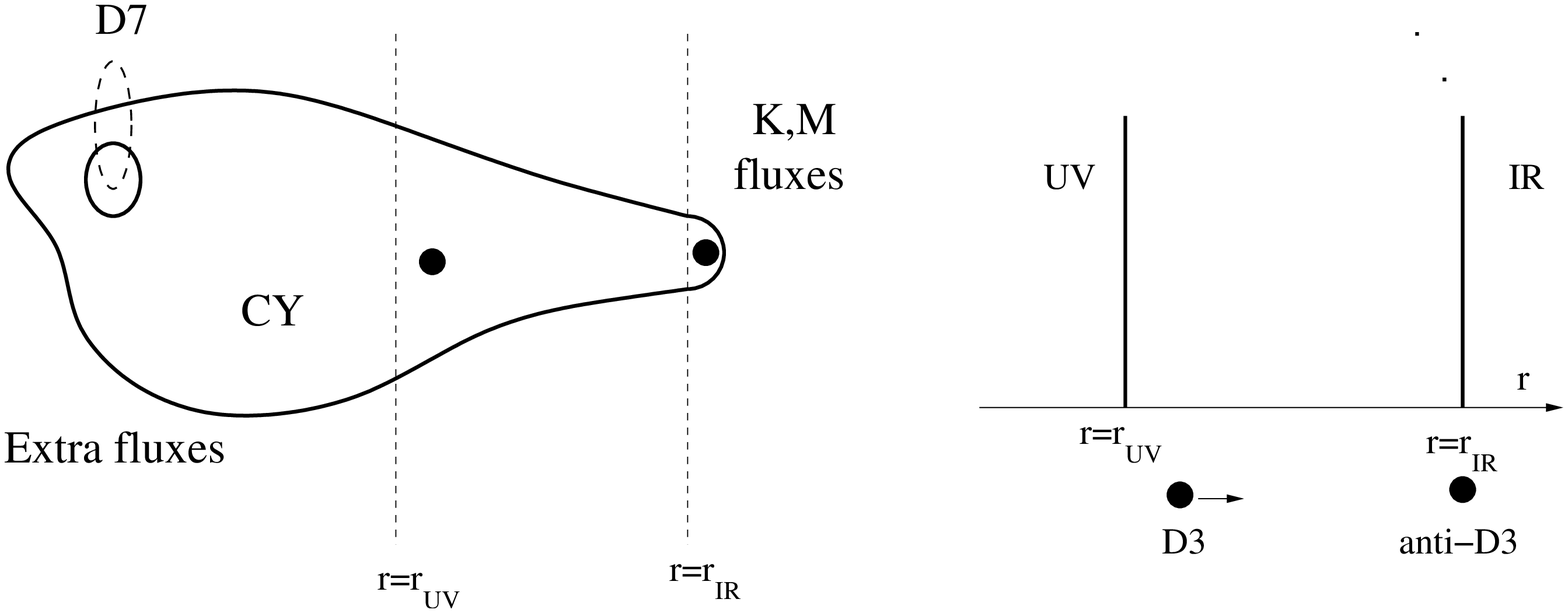}}
Fig. 1: {\it The Calabi-Yau compactification with a throat
and its corresponding interpretation in terms of a simplified
RSI model. 
}
\endinsert

As a specific example, let us discuss the KKLMMT model \KKLMMT\ for $D$-brane 
inflation\footnote{$^1$}{The KKLMMT model was proposed in order
to overcome some of the problems of the $D$-brane inflation
scenario in flat space but it is still plagued
by the familiar $\eta$ problem of F-term
inflation in supergravity (the brane position appears in the K\"ahler
potential and the stabilization of the volume produces a mass 
for the inflaton of order $H$).
Various ways to solve this problem, with or without fine tuning, 
have been proposed in ref.~\KKLMMT\ and in
subsequent works on the subject \KKLMMTrefs .
Here we take the conservative attitude of assuming that this problem
has been solved and look for the presence of massless fields during the 
inflationary period.}. 
The KKLLMT model is based on a compactification of the type IIB string
theory on a Calabi-Yau manifold in the presence of orientifolds and
fluxes for the antisymmetric tensors. As advocated in refs.~\refs{\GKP,\KKLT},
the fluxes can stabilize all the complex structure moduli of the 
compactification (the shape of the Calabi-Yau). The K\"ahler moduli
(the volume) can be then stabilized by non-perturbative
superpotentials generated by additional $D$7 branes (or euclidean instantonic
$D$3-branes) naturally present in such compactifications \refs{\KKLT,\DD}.
The fluxes also induce a warp factor in the metric. 
The KKLMMT scenario corresponds to a Calabi-Yau geometry that develops
a long warped AdS-like throat. 
The model may be thought as 
a  stringy realization of the Randall-Sundrum model
(RSI) \rsI , where the infra-red 
brane has been effectively regularized by an infra-red geometry and 
the ultra-violet brane has been replaced by the Calabi-Yau geometry
(see Fig. 1).

The entire physics of inflation takes place in the throat region and
it is mostly insensitive to the details of the ultra-violet Calabi-Yau
region. In the KKLMMT an anti $D$3-brane is sitting at the infra-red end 
 of the throat and a $D$3-brane is moving toward it. In the absence of 
anti branes, 
the $D$3 is a supersymmetric BPS object and it feels no force \KS\ . 
The only source of supersymmetry breaking in the compactification
is the inclusion of the anti $D$3-brane.
The insertion of a distant anti $D$3-brane just 
produces a very flat potential \KKLMMT\ for the scalar field corresponding 
to the position of the
$D$3-brane, which plays the role of the inflaton field. The potential
for the canonically normalized inflaton $\phi$ is
\eqn\pot{V(\phi)= T_3 a_0^4 \left (1-{C\over \phi^4}\right ) ,}
where $T_3$ is the tension of a $D$3 brane, $a_0$ is the redshift
factor at the bottom of the throat and $C$ is a numerical constant whose
value can be found in ref.~\KKLMMT . If we
ignore the order $H$ mass that the inflaton will acquire due to
the K\"ahler potential \KKLMMT , the condition of slow roll and the constraints from
the density perturbations can be satisfied with
reasonable values of the parameters\footnote{$^2$}{In ref.~\KKLMMT\
the integer fluxes $M,K$ are taken 
${\cal O}(10)$, $T_3/M_p^4\sim 10^{-3}$  and
the  warping is mild $a_0\sim 10^{-4}$. The Hubble scale is consequently of 
typical order $10^{9}$ GeV.}.

For our purposes, we notice that the KS solution \KS\ has an 
$SU(2)\times SU(2)$ isometry group. 
During its motion toward the
infra-red, the $D$3-brane can fluctuate in the internal angular directions.
Since during the inflationary period, the two brane set-up 
preserves the $SU(2)\times SU(2)$ symmetry, 
the scalar fields associated with the angular positions are almost
massless and lighter than the Hubble rate. Furthermore, being angles, 
they do not get masses from the volume stabilization
mechanism. In the K\"ahler potential
of four-dimensional supergravity the volume modulus $\rho$ always appears in 
the
combination $\rho+\bar\rho - k(\phi_i,\bar\phi_i)$ \GD , 
where $\phi_i$ collectively
denote the position of the branes in the six internal directions,
and $k$ is the K\"ahler potential for the geometry.
This coupling generates a mass for the fluctuations $\phi_i$.
However, since the geometry has isometries, 
$k(\phi,\bar\phi)$ does not depend on some of the angles. In this way, the 
form of potential for the angular
fluctuations is not affected by the stabilization mechanism and leaves
some angles much lighter than the Hubble rate during inflation.
At the end of inflation, the $D$3-brane comes close to the anti $D$3-brane.
At stringy distances, we cannot ignore the fields coming from
open strings connecting the branes. In particular, in a generic 
brane-antibrane system, the open string ground state is a tachyon $T$ 
with mass of order $M_s$,
the string scale. Since the annihilation takes
place at the bottom of the throat, the tachyonic mass is redshifted to
$m_T^2\sim -a_0^2M_s^2$.  
We can write 
a toy-model potential for the inflaton $\phi$, the tachyon $T$ and the
angular positions $\Theta_i$  which captures the salient 
features of the tachyonic condensation when the brane and anti-brane
are close to each other
\eqn\pot{V(\phi,\Theta_i,T)=T_3 a_0^4 \left (1-{C\over \phi^4}\right )+
T^2(\phi^2-a_0^2M_s^2+\tilde V(\Theta_i))+\cdots ,} 
where the ellipses stand for higher-order power terms in the field $T$.
In particular, the form of the potential accounts for
the  $SU(2)\times SU(2)$ sources of breaking when the two branes are
close. 
Notice that this
scenario is similar to that discussed in sect.~\S3\  in the context of 
hybrid inflation. As long as
$\phi$ is much larger that the (redshifted) string scale, 
$T$ has a large positive 
mass, its vacuum expectation value $\langle T\rangle$ vanishes
 and all the terms in the potential involving $T$ can be ignored.
The inflationary potential is then independent of the angles $\Theta_i$. 
At stringy distances, $T$ becomes tachyonic
and triggers the brane-antibrane annihilation. When $T$ condenses,
all the fluctuation fields on the
branes acquire masses roughly of order of the redshifted string scale. 
Taking into account that $a_0\sim 10^{-4}$ at the bottom of 
the throat, these masses are ${\cal O}(a_0M_s)$ and much  larger than $H\sim a_0^2M_s$. 
All these massive fields will eventually decay. 

There are two effects that break the $SU(2)\times SU(2)$ invariance and 
may prevent the angular fluctuations of the brane 
from being exactly massless during inflation. 
There is a small source of breaking
of the  $SU(2)\times SU(2)$ symmetry due to the presence of the anti
$D$3-brane
in the infra-red, which sits at a specific point on the three-sphere.
This explicit breaking is though suppressed when the 
the distance between the two branes during the inflatonary
period is sizeable.
The second
effect is caused by the moduli stabilization that
requires fluxes in the ultra-violet region typically breaking
$SU(2)\times SU(2)$; it is well known indeed that a Calabi-Yau manifold
has no isometries at all. Even this second effect is suppressed by the
distance. An estimate based on the arguments in appendix A.2 of ref.~\giant\
shows that the typical mass induces by the $SU(2)\times SU(2)$ breaking
scales as $a^8(\phi)$~\footnote{$^3$}{Actually, the smallest breaking effect
could be of order $a^5(\phi)$ with a generic Calabi-Yau. As in ref.~\giant\
we can however choose a manifold with a discrete $Z_2$ symmetry
that forbids this effect.}, where $a(\phi)$ is 
the warp factor evaluated at the position $\phi$ of the brane. 
On the other hand, using formulae in appendix C of  ref.~\KKLMMT ,
we can easily evaluate $a(\phi)\sim a_0^{2/3}$ when there are
about 60 $e$-folds till the end of inflation.
Therefore, the mass of the angular variables
$m_\Theta^2\sim a_0^{16/3}$ is suppressed compared to $H\sim a_0^4$
by powers of the redshift factor $a_0\sim 10^{-4}$. 

We finish this section by commenting on the dual interpretation of these
massless fields. As familiar from the AdS/CFT correspondence and the
holographic interpretation of the RS model, the throat part of the 
compactification  can be effectively replaced by a strongly interacting
gauge theory coupled to the four-dimensional gravity. In this picture,
the $D$3 brane position corresponds to a flat direction in the moduli space
of vacua of the gauge theory.
The global symmetry $SU(2)\times SU(2)$ is spontaneously broken along
this flat direction\footnote{$^4$}{The KS model with
K and M fluxes is dual to a $SU(KM+M)\times SU(KM)$ $N=1$ supersymmetric
theory with pairs of bifundamental fields $A_i$ and $B_i$, $i=1,2$,
each one rotated by a global $SU(2)$ symmetry, with an $SU(2)\times SU(2)$
invariant superpotential. The same compactification with a $D$3 brane
inserted can be represented by a gauge group
 $SU(KM+M+1)\times SU(KM+1)$ broken to $SU(KM+M)\times SU(KM)$
along a flat direction where $A,B$ acquire a VEV breaking the 
global symmetry $SU(2)\times SU(2)$.}.  
The angular
positions of the brane are identified  with the massless Goldstone bosons of 
the spontaneously  broken global symmetry.

\newsec{Applications of heavy particle production from inflation}
In sect.~\S2\ we have estimated the 
number density of heavy particles generated during inflation in the
case in which the large masses are acquired after inflation. The
generation mechanism does not directly involve the particle mass
and it can be very efficient.  We now proceed to 
describe some applications of our previous
findings. 

\subsec{Superheavy dark matter}
The case for dark nonbaryonic matter in the universe is today stronger than
ever.  The observed large-scale structure suggests that dark
matter (DM) accounts for about 30\% of the critical mass density of the
universe $\rho_C= 1.88\times 10^{-29}$ g cm$^{-3}$ \turner .
Despite this compelling evidence, the nature of the DM is still unknown. Some
fundamental physics beyond the Standard Model (SM) is certainly required to
account for the cold and slowly moving particles, $\chi$, composing the bulk
of the nonbaryonic dark matter.  
The most familiar assumption is that dark matter is a thermal relic, {\it
i.e.,} it was initially in chemical equilibrium in the early universe
and its present-day abundance is inversely proportional 
to the annihilation cross section.
Since the  largest possible annihilation cross section
is roughly $M^{-2}$, one expects a maximum mass for a thermal DM particle, 
which turns out to
be a few hundred TeV \griestkam .
While a thermal origin for DM paricles 
is the most common assumption, it is not the
only possibility.  
DM particles may be produced independently of any
microphysics process.  
It has been pointed out that DM particles might have
never experienced local chemical equilibrium during the evolution of the
universe and that their mass may be in the range $10^{12}$ to $10^{19}$ GeV,
much larger than the mass of thermal relics 
\refs{\ckra,\ckrb,\ckrc,\ckrd,\ckre}. 
These superheavy
DM particles have been called WIMPZILLAs \ckrd .  
Independently of the
production mechanism, the fraction of the total energy density
of the Universe in superheavy DM particles today is given by
\eqn\abund{\Omega_\chi  = \Omega_R \,
\left({T_r\over T_0}\right)\, \left.{\rho_\chi\over\rho_\phi}\right|_*.}
Here, $\Omega_R h^2 =4.3  \times 10^{-5}$ is the fraction of critical
energy density in radiation today, $T_0$ is the present temperature of
radiation, and the subscript indicates the epoch at which
heavy particles are generated. The present abundance of nonthermal 
DM particles is,
as expected, independent of the cross section \refs{\ckra,\ckrb}, and one can
easily verify that, if it amounts to
$\Omega_\chi\sim 1$, nonequilibrium during the evolution of the
universe is automatic.  

Several scenarios for superheavy DM particles have been developed,
which involve
production during different stages of the evolution of the universe. They may
be generated in the transition between an inflationary and a matter-dominated
(or radiation-dominated) universe due to the ``nonadiabatic'' expansion of the
background spacetime acting on the vacuum quantum fluctuations. This mechanism
was studied in details in Refs. \refs{\ckra,\kuzmin}, in the case of chaotic
inflation. The distinguishing feature of this mechanism is the capability of
generating particles with mass of the order of the inflaton mass (usually much
larger than the reheating temperature, but at most of the 
order of the Hubble rate during inflation) 
even when the particles only interact
extremely weakly (or not at all) with other particles, and do not couple to the
inflaton. Superheavy DM particles 
may be also created during bubble collisions if inflation
is completed through a first-order phase transition \mr ; at the
preheating stage after the end of inflation with masses up to 
$10^{15}$GeV if they are bosons \Kolba , and up to the
Planck scale if they are fermions \Giudicea ; or during the
reheating stage after inflation \ckrc\ with masses which may be as large
as $2\times 10^3$ times the reheat temperature. 
In addition, particles as heavy as the inflaton can be
efficiently produced perturbatively, both in inflaton decay
\ad\ and during thermalization of
energetic inflaton decay products~\da .

We can use the production mechanism described in sect,~\S2
and imagine that superheavy DM particles are massless during inflation and
get their large mass right after inflation. In such a case, 
the present-day abundance is readily computed 
using eqs.~\rr\ and
\abund\  and we find $\Omega_\chi h^2 =0.1$ for
\eqn
\omega
{M\simeq \left( {\epsilon \over 10^{-2}}\right)^{1/2}
\left( {{\rm TeV}\over T_r}\right)^{1/2} 4\times 10^{12}~{\rm GeV}.}
The request that the DM particles are not produced by thermal processes
($M\ga T_r$), implies that 
\eqn\pippa
{T_r \la \left( {\epsilon \over 10^{-2}}\right)^{1/3}
3\times 10^{9}~{\rm GeV}.}
Therefore the proposed mechanism can account for DM in a large range
of $T_r$ which, in particular, can be compatible with an acceptable 
regeneration of thermal gravitinos. The upper bound on $T_r$
reflects the extreme efficiency
of the mechanism described in this paper 
in creating heavy particles from inflation.

\subsec{Baryogenesis}

Another natural application for the proposed mechanism of heavy-particle
production is baryogenesis. Since baryon-number violation 
(or lepton-number violation, if sphalerons
are active) is a necessary ingredient of baryogenesis,
the relevant dynamics generally takes place at very short 
distances and involves very heavy particles. This is often at odds
with the request of sufficiently low $T_r$, coming from cosmological
considerations on unwanted relics like, for instance, gravitinos or
moduli (for a recent analysis, see ref.~\kawa ). 
A mechanism for producing particles with masses much larger
than $T_r$ could resolve this conflict.

The situation is well illustrated by leptogenesis \lepto , 
one of the most concrete
mechanisms to successfully reproduce the observed cosmic baryon asymmetry.
Therefore we will focus on this case, although our discussion can be
extended also to
the case in which baryogenesis occurs through the decay of some GUT field,
nearly massless during inflation. 

Thermal leptogenesis implies a lower bound on $T_r$, which depends on
the right-handed neutrino mass, but it is always larger than 
about $10^9$~GeV \leptonoi . 
If we want to consider lower values of $T_r$, we should rely either 
on variations of leptogenesis (soft leptogenesis~\softl,
resonant leptogenesis~\pilaf, leptogenesis with sneutrino 
dominance~\refs{\sndom,\allah})
or
on non-thermal production of the right-handed neutrinos $N$.
One possibility
is to generate $N$ at reheating by direct
inflaton decay \infa . Alternatively, the right-handed neutrinos
could be produced during the
reheating stage when temperatures much larger than $T_r$ are actually
achieved \ckrc . If a preheating stage occurs, 
when large inflaton oscillations decay non-perturbatively,
than a non-thermal density of heavy
fermions can be generated \Giudicea . Here we want to study if the mechanism
proposed in this paper could also create a sufficiently large population
of right-handed neutrinos.

Since $N$ are fermions, we cannot take direct advantage of 
fluctuations generated at inflation. The estimate given in eq.~\pip\
shows that their density is too small to account for the baryon asymmetry.
However, there are still some interesting possibilities for leptogenesis.

First, we can consider the scalar field $\phi$ whose 
lepton-number violating vev gives
mass to the right-handed neutrinos though the coupling $\phi N^{cT}N$.
If $\phi$ is massless during inflation and acquires a mass $M_\phi$
right after, it can play
the role of the field $\chi$ described in sect.~\S2. In 
particular, this could happen if $\langle \phi \rangle$ is the dominant
source of lepton violation during inflation; then 
the associated pseudo-Goldstone
boson could obtain large fluctuations.
The large $\phi$
number density given by eq.~\ratio , after decay, will be transferred to $N$.
Assuming that $\phi$ dominantly decays into $N$, the baryon
asymmetry is given by
\eqn\bar
{{n_B \over s} =a~ \epsilon_{CP} ~{M_\phi T_r \over 4\pi \epsilon M_p^2}.}
Here $a=0.35$ (both in the Standard Model and in minimal supersymmetry)
relates the $B-L$ to the $B$ asymmetry through sphaleron interactions.
For hierarchical neutrino masses, the CP asymmetry in $N$ decays 
($\epsilon_{CP}$) has the upper bound \bound
\eqn\cp
{\epsilon_{CP}< {3M_N m_\nu^{\rm atm} \over 16 \pi v^2 }=
3\times 10^{-7} {M_N \over 10^{10}~{\rm GeV}},}
where $M_N$ is the mass of the lightest right-handed 
neutrino\footnote{$^4$}{This bound can be relaxed if the neutrino-mass
hierarchy is reduced~\strum .}.
Requiring that eq.~\bar\ reproduces the observed value
$n_B/s =8.7 \times 10^{-11}$ 
and using eq.~\cp , we obtain the constraint
\eqn\rutt
{{T_r \over M_N} > \left( {10^{15}~{\rm GeV}\over M_N}\right)^2
\left( {10^{18}~{\rm GeV}\over M_\phi }\right)
\left( {\epsilon \over 10^{-2}}\right) \times 10^{-4}.}
This shows that, for heavy $N$, it is possible to generate the
correct baryon asymmetry even if $T_r$ is several orders of
magnitude smaller than $M_N$. This region of parameters 
($T_r \ll M_N$) is not
accessible to the standard thermal leptogenesis. The proposed mechanism
is most efficient when $M_N$ is large, close to the GUT scale. Indeed,
absence of thermal processes ($M_N\ga T_r$) gives a lower bound on
the right-handed neutrino mass
\eqn\rutb
{M_N\ga\left( {10^{18}~{\rm GeV}\over M_\phi }\right)^{1/2}
 \left({\epsilon \over 10^{-2}}\right)^{1/2} \times 10^{13}~{\rm GeV}.}

Supersymmetry offers another possibility. 
The right-handed sneutrino 
could be the
source of leptogenesis and,
if its mass term is suppressed during inflation,
its density before decay will be given
by eq.~\ratio . 
The same argument that lead us to eqs.~\rutt\ and \rutb\ now gives us
the constraints
\eqn\flip
{T_r>  {\epsilon \over 10^{-2}}\left( {10^{15}~{\rm GeV}\over M_N}\right)^2
2\times 10^{14}~{\rm GeV}.}
\eqn\flop
{M_N \ga
 \left({\epsilon \over 10^{-2}}\right)^{1/3}6\times 10^{14}~{\rm GeV}.} 
These bounds are more stringent than in the previous case, where a
larger density of right-handed neutrinos is achieved by considering
very large $\phi$ masses.

\newsec{Conclusions and discussion}

In this paper we have demonstrated 
 how a primordial epoch of inflation might 
be associated to  the generation of very heavy particles. 
The mechanism is rather simple once it is realized that heavy
states might have been either exactly or nearly massless during inflation.  
Long-wavelength fluctuations of such fields are generated during inflation
with an almost scale-invariant spectrum 
and might provide  the largest contribution to their energy density at 
late 
times. This mechanism is 
present not only 
in effective quantum field theories, but also in  brane-world scenarios 
inspired by  string theory where the lightness of 
would-be heavy states during inflation is guaranteed by 
exact or approximate isometries of the background geometry. We have 
described applications of our findings to 
superheavy dark matter and leptogenesis.

There are other interesting issues which would deserve further and careful
investigation. First of all, the mechanism discussed in this paper
might provide a natural way to implement the curvaton mechanism 
to generate 
cosmological perturbations. It  has been proposed that the field 
responsible for the
observed cosmological perturbations is some
 `curvaton' 'field different from the inflaton \curvaton . During
inflation, the curvaton energy density is negligible and isocurvature
perturbations with a flat spectrum are produced in the curvaton field.
After the end of inflation, the curvaton field oscillates during some
radiation-dominated era, causing its energy density to grow and thereby
converting the initial isocurvature into curvature perturbation.  To be
operative the curvaton mechanism requires the mass of the curvaton to be
lighter than the Hubble rate during inflation and decay must occur after
inflation. As we have shown, these requirements are satisfied, for 
instance, by brane models
where  the distance between a $D$-brane and an anti $D$-brane 
plays the role of the inflaton. Brane fluctuations
transverse to the mutual distance give rise to
almost massless fields during inflation, which become heavy and decay
once inflation is terminated. 
The curvaton acts
as a pseudo-Goldstone boson whose dynamics has been thoroughly studied in 
ref.~\pngb . 

Our findings might also be relevant for the dynamics of
reheating after inflation. One can envisage two extreme cases. Suppose
that the decay rate of the inflaton field into ordinary matter is highly 
suppressed. If so, populating the Universe  with known particles might occur
through $\chi$-states which are efficiently generated during inflation
and very massive after inflation. On the contrary, if the 
inflaton couplings to ordinary matter are of order unity, defrosting the 
Universe after inflation 
is very efficient and might lead  to  high reheating   temperatures
$T_r$, possibly in conflict with the gravitino bound \kawa . This
drawback could be avoided if there is a subsequent release
of entropy caused by the decay of heavy $\chi$-states abundantly
produced during inflation.



\listrefs

\bye

\end